# High temperature superconductivity of quaternary hydrides $XM_3Be_4H_{32}$ (X, M = Ca, Sr, Ba, Y, La, Ac, Th) under moderate pressure

Wendi Zhao[1], Defang Duan[1]*, Decheng An[1], Qiwen Jiang[1], Zhengtao Liu[1], Tiancheng Ma, Zihao Huo[1], Jianhui Du[1] and Tian Cui*[1,2]

[1]State Key Laboratory of Superhard Materials, College of Physics, Jilin University, Changchun 130012, China

[2]Institute of High Pressure Physics, School of Physical Science and Technology, Ningbo University, Ningbo 315211, China

**Abstract**

The compressed hydrogen-rich compounds have received extensive attention as promising candidates for room temperature superconductivity, however, the high pressure required to stabilize such materials hinders their wide practical application. In order to search for potential superconducting hydrides that are stable at low pressures, we have investigated the crystal structures and properties of quaternary hydrides, $XM_3Be_4H_{32}$ (X, M = Ca, Sr, Ba, Y, La, Ac, Th) based on the first-principles calculations. We identified nine dynamically stable compounds at moderate pressure of 20 GPa. Strikingly, their superconducting transition temperatures are much higher than that of liquid nitrogen, especially $CaTh_3Be_4H_{32}$ (124 K at 5 GPa), $ThLa_3Be_4H_{32}$ (134 K at 10 GPa), $LaAc_3Be_4H_{32}$ (135 K at 20 GPa) and $AcLa_3Be_4H_{32}$ (153 K at 20 GPa) exhibit outstanding superconductivity at mild pressures. Metal atoms acting as pre-compressors donate abundant electrons to hydrogen, weakening the H-H covalent bond and thus facilitating the metallization of the hydrogen sublattice. At the same time, the appropriate combination of metal elements with different ionic radius and electronegativity can effectively tune the electronic structure near the Fermi level and improve the superconductivity. These findings fully reveal the great promise of hosting high-temperature superconductivity of quaternary hydrides at moderate pressures and will further promote related exploration.

## 1. Introduction

Compressed hydrogen-rich compounds are considered as one of the best candidates for high-temperature superconductors and have received considerable attention as the frontiers of physics and material science. In recent years, the breakthrough in the study of binary superconducting hydrides is exciting, and a considerable number of them have been confirmed to have excellent superconductivity, such as $H_3S$[1-3], $LaH_{10}$[4-9], $YH_6$[10, 11], $YH_9$[11, 12], $CaH_6$[13], etc. Multi-component hydrides with richer chemical compositions and crystal structures offer a unique opportunity to further optimize the properties of materials, hosting great prospects for exhibiting more novel properties, such as $Li_2MgH_{16}$[14], $Li_2NaH_{17}$[15], $LiNa_3H_{23}$[15], etc., which were theoretically reported to host room-temperature superconductivity. Undoubtedly, superconducting hydrides have established increasingly high critical temperature records, which is very exciting[16, 17]. From the reported results, however, hydrides with excellent superconductivity are usually stable above 150 GPa[18]. Although this is much lower than the stable pressure of metal hydrogen[19], it is still far from the goal of wide practical applications of superconducting hydrides. Hence, the next key scientific issue is to lower the stable pressure of superconducting hydrides, and the ultimate goal is to achieve room temperature superconductivity at significantly low pressure or even ambient pressure.

Doping represents one of the most promising avenues to tune the properties of materials, which has been reflected in the study of multi-component superconducting hydrides. In general, for metal hydrides, metal atoms can donate abundant electrons to hydrogen, thereby weakening the H-H strong covalent bond, which may drive the formation of novel hydrogen motifs. For example, Li atoms doped into $MgH_{16}$, which is rich in $H_2$ molecular units, drive the $H_2$ units to expand to form clathrate covalent lattice. The predicted $T_c$ value of $Li_2MgH_{16}$ at 250 GPa is as high as 473 K[14]. Furthermore, the substitutional-doping is also expected to lower the stable pressure of superconducting hydrides[20]. For example, the dynamically stable critical pressure of $YLu_3H_{24}$[21] is lower than that of $MgH_6$[22] and $YH_6$[10] with the same hydrogen motif, exhibiting high $T_c$ up to 288 K at 110 GPa. Recently, theory predicts that alloy sublattices formed by small radius non-metallic elements (e.g., Be, B, C, etc.[23-27]) with hydrogen are more easily stabilized at moderate pressures than pure hydrogen lattices. For example, in $LaBeH_8$[23], the metal La produces effective chemical pre-compression on the alloy backbone formed by Be and H, resulting in $LaBeH_8$ being able to maintain dynamic stability at as low as 20 GPa and exhibiting a high $T_c$ of 185 K. Recently, $LaBeH_8$ has been successfully synthesized by experiment, and the $T_c$ value measured at 80 GPa reaches 110 K[28]. These excellent properties are unprecedented in binary hydrides, indicating the potential advantages of multi-component hydrides in coordinating the advantages of different elements to induce novel properties. Notably, the construction of complete phase diagrams for multi-element (e.g, quaternary and pentabasic) superhydrides is very challenging, which not only incurs high computational costs but also goes beyond current computational capabilities. Therefore, performing substitution-doping of a priori hydrides with excellent performance is currently an effective strategy to accelerate the screening of multi-component phases.

Here, inspired by the recently synthesized rocksalt-like $LaBeH_8$[28], we constructed novel quaternary superhydrides $XM_3Be_4H_{32}$ (X,M = Ca, Sr, Ba, Y, La, Ac, Th) by substitutional-doping the supercell of $LaBeH_8$, and systematically investigated their structural stability and superconducting properties. Relying on first-principles calculations combined with high-throughput screening, we identified nine dynamically stable high-temperature superconducting quaternary hydrides at 20 GPa: $AcTh_3Be_4H_{32}$, $LaTh_3Be_4H_{32}$, $BaTh_3Be_4H_{32}$, $SrTh_3Be_4H_{32}$, $CaTh_3Be_4H_{32}$, $ThAc_3Be_4H_{32}$, $ThLa_3Be_4H_{32}$, $LaAc_3Be_4H_{32}$ and $AcLa_3Be_4H_{32}$. Strikingly, their maximum $T_c$ values are much higher than the liquid nitrogen temperature, especially $CaTh_3Be_4H_{32}$, $ThLa_3Be_4H_{32}$, $LaAc_3Be_4H_{32}$ and $AcLa_3Be_4H_{32}$ exhibit outstanding superconductivity at moderate pressures, and thus their superconducting figure of merit S exceed 3, hosting the highest values among quaternary hydrides. The excellent superconductivity is attributed to the effective adjustment of the H-H bond length by different metals as pre-compressors, which induces high H-driven electron density of states at the Fermi level and enhances the electron-phonon coupling. These findings highlight the great potential of multi-hydrides to achieve high-temperature superconductivity under moderate pressure and even ambient pressure, and provide key insights for further exploration.

2. **Computational methods**

We performed structural relaxations and electronic properties calculations of selected quaternary phases within the framework of density functional theory (DFT), as implemented in Vienna ab initio simulation package (VASP) code[29]. The ion-electron interactions part was implemented with the projector augmented wave (PAW) method[30, 31]. The exchange-correlation functional was described using the Perdew-Burke-Ernzerhof (PBE) parametrization within the generalized gradient approximation (GGA)[32]. The cutoff energy was chosen to be 1000 eV and the Monkhorst-Pack $k$-mesh with grid spacing of $2\pi \times 0.02$ Å$^{-1}$ was adopted to ensure that the enthalpy calculations were converged to within 1 meV/atom. The crystal orbital Hamiltonian population (COHP)and its integral (ICOHP) were calculated using the LOBSTER code[33, 34]. The electron-phonon coupling (EPC) calculations were carried out using the Quantum ESPRESSO package[35]. Ultra-soft pseudopotentials were selected with a kinetic energy cut-off of 80 Ry. For the considered $XM_3Be_4H_{32}$ compounds, the *k*-meshes of $2\pi \times 0.02$ Å$^{-1}$ and *q*-meshes of $2\pi \times 0.05$ Å$^{-1}$ were used to accurately calculate the electron-phonon coupling. The Allen-Dynes modified McMillan equation[36] is used to calculate the superconducting transition temperature ($T_c$).

## 3. Results and Discussion

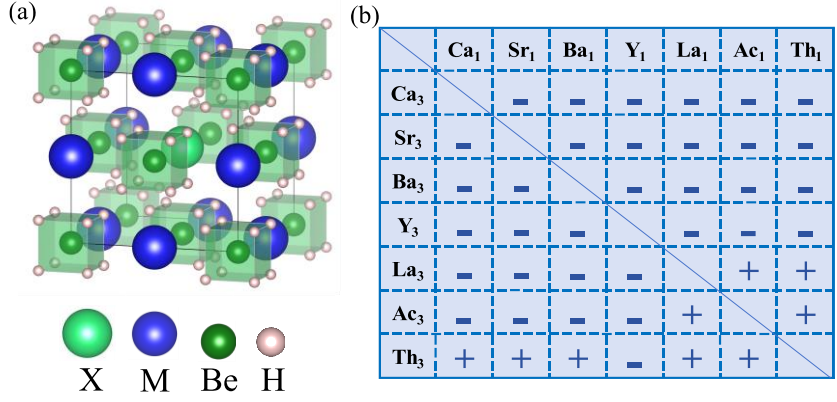

Fig. 1 (a) Structural motifs of the quaternary hydrides $XM_3Be_4H_{32}$ with *Pm*-3*m* symmetry, the number of metal atom occupying the X-site in the cell is 1 and the number of metal atom occupying the M-site is 3. (b) Results of high-throughput calculations of phonon dispersion curves for $XM_3Be_4H_{32}$ compounds at 20 GPa. Phases that are dynamically unstable from phonon calculations are labeled with '−', and stable phases are labeled with '+'.

We constructed quaternary hydrides $XM_3Be_4H_{32}$ (X, Y = Ca, Sr, Ba, Y, La, Ac, Th) with *Pm*-3*m* symmetry composed of two different metal atoms pre-compressed $BeH_8$ units. Metal atoms X and M occupy the 1b (0.50, 0.50, 0.50) and 3d (0.50, 0.00, 0.00) Wyckoff positions, respectively. Be atoms occupy 1a (0.00, 0.00, 0.00) and 3c (0.00, 0.50, 0.50) Wyckoff positions, which are located in the center of $H_8$ cubic units. The reasonable choice of metal atoms X and M as pre-compressors refers to previous theoretical reports[23, 37]. We performed high-throughput calculations for the $XM_3Be_4H_{32}$ compounds. In total of 42 compounds were relaxed at 20 GPa (see Figure 1), and their dynamic stability was further evaluated via phonon dispersion calculations. As a result, nine quaternary hydrides $AcTh_3Be_4H_{32}$, $LaTh_3Be_4H_{32}$, $BaTh_3Be_4H_{32}$, $SrTh_3Be_4H_{32}$, $CaTh_3Be_4H_{32}$, $ThAc_3Be_4H_{32}$, $ThLa_3Be_4H_{32}$, $LaAc_3Be_4H_{32}$ and $AcLa_3Be_4H_{32}$ are dynamically stable at 20 GPa (see Figure S1), and also among them, four hydrides $CaTh_3Be_4H_{32}$, $LaTh_3Be_4H_{32}$, $SrTh_3Be_4H_{32}$, $ThAc_3Be_4H_{32}$ could be dynamically stable at 5 GPa. At the same time, these compounds have significant energy advantages above 50 GPa compared to the possible decomposition enthalpies (see Figure S2). Notably, in terms of stable chemical fractions, they contain heavy rare earth elements and actinides with similar electronegativity and atomic radii, revealing the importance of good size match for the metal ions with one another, as well as for the Be-H backbone encapsulating them. In particular, some heavy elements with empty *f*-shells (e.g., La, Ac, Th) are excellent pre-compressors for the XM$_3$Be$_4$H$_{32}$ structural motif, which is also reflected in the previously reported XBeH$_8$ (X = La, Ac, Th)[38, 39]. It is worth noting that previous studies have shown that an increase in the occupation of *f*-shell electrons will inhibit the superconductivity of hydrides[21], and therefore elements with occupied *f*-shell electrons (e.g., Ce, Yb, etc.) are not considered in this work.

We investigated the electronic properties of these dynamically stable quaternary hydrides. The absence of electronic localization between metal and H atoms confirms ionic interactions (see Figure S3). Note that the ionic properties differ slightly between the metal atoms and the BeH$_8$ unit, which may depend on their different valence states and electronegativities. The average H-H bond lengths in these compounds ranged from 1.70 ~ 1.78 Å, which is much greater than the 0.72 Å for H$_2$ molecules at ambient pressure. However, the electronic localization function (ELF) values within the H$_8$ unit are between 0.55 ~ 0.75, suggestive of weakly bound hydrogen bonding motifs. Moreover, although the H - H distance in the H$_8$ unit is longer than that between the H$_8$ units, its ELF value is slightly larger than the ELF value between the H$_8$ units. For example, for CaTh$_3$Be$_4$H$_{32}$, the average H-H bond length (ELF value) is 1.74 (0.75) within the H$_8$ unit and 1.68 (0.56) between the H$_8$ units. In fact, this is related to the distribution of electron localization in the interstitial region of the metal sublattice. Significant electron localization is generated in the octahedral gap region of the isolated metal sublattice, which matches well with the positions of the H$_8$ units, thereby stabilizing the hydrogen bonding motifs (see Figure S4). Miao et al. proposed the chemical template theory to reveal this point well[40], that is, electrons in the interstitial region of the metal sublattice occupy quantum orbitals here and form a template assisting the assembly of the hydrogen sublattice, thus maintaining the weak covalent interaction between the long H-H distance. Importantly, the chemical template effect minimizes the total energy of the metal and hydrogen lattices, providing a strong driving force for structural stability.

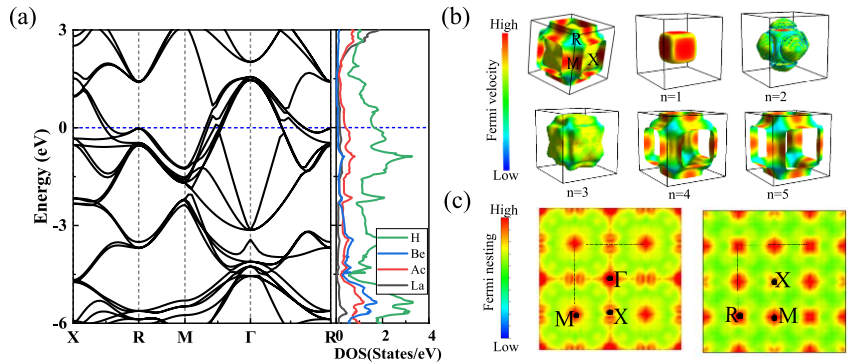

Fig. 2 (a) Calculated electronic band structures and projected density of states (PDOS) for LaAc$_3$Be$_4$H$_{32}$ at 20 GPa. The Fermi level is set to zero. (b) 3D Fermi surface sheets of LaAc$_3$Be$_4$H$_{32}$ in Brillouin zone and the corresponding FS sheets for the five bands (n = 1–5). (c) 2D Fermi surface nesting of $LaAc_3Be_4H_{32}$ in Brillouin zone.

As mentioned above, different combinations of metal atoms acting as pre-compressors closely influence the atomic bonding, which will further tune the electronic structure. The calculated band structure and electronic density of states of these stable hydrides have some common characteristics (see Figure S5). They all exhibit rigid band behavior, and multiple steep electronic bands cross the Fermi energy level to form "electron pockets" around the M point. The electronic bands associated with metal atoms are mainly localized above the Fermi level, and their overlap with H-related bands reveals strong hybridization interactions. Importantly, hydrogen contributes greatly to the DOS at the Fermi level, which is an important prerequisite for high $T_c$ conventional superconductivity in hydrides. Notably, there are two almost parallel bands crossing the Fermi level along the X-R directions. The appearance of parallel bands in the electronic band structure implies the possibility of Fermi surface nesting. The strength of Fermi surface nesting depends on the shape of the Fermi sheet. We calculated the Fermi surface and Fermi surface nesting of these nine quaternary structures in Brillouin zone (see Figure S6-7). Note that the nesting of $ThAc_3Be_4H_{32}$, $ThLa_3Be_4H_{32}$, $LaAc_3Be_4H_{32}$ and $AcLa_3Be_4H_{32}$ at X, R, and M points is very obvious, which is attributed to the fact that the pre-compressors composed of different metal elements effectively adjusts the Fermi energy, bringing the flat electronic bands closer to the Fermi levels. For example, for $LaAc_3Be_4H_{32}$, there are multiple Fermi sheets parallel to each other in the Brillouin zone, especially the 4th and 5th ones, implying that nesting may occur (see Figure 2). Thus, strong Fermi surface nesting occurs near the X and M points along the Γ-X and Γ-M directions, respectively, which is associated with parallel Fermi surfaces in these directions. In addition, the Γ-R direction has smaller Fermi velocities and thus also exhibits strong Fermi surface nesting near the R point[41]. Importantly, the strong electron-phonon coupling of the nested electronic states on the Fermi surface plays an important role in improving superconductivity.

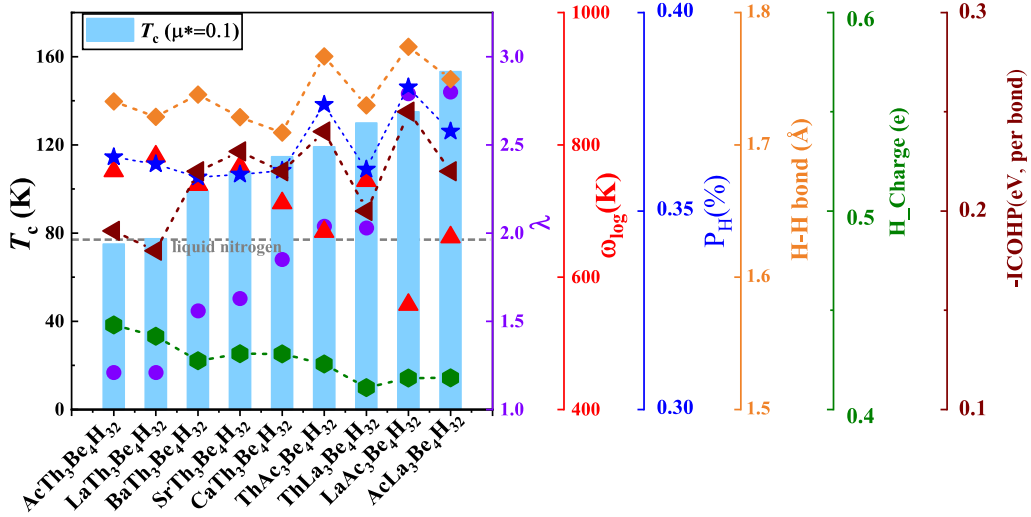

Fig. 3 The $T_c$ and related parameters of different $XM_3Be_4H_{32}$ compounds at 20 GPa. The obtained $T_c$ using the Allen-Dynes modified McMillan equation ($\mu^* = 0.1$), EPC parameter ($\lambda$), The logarithmic average phonon frequency $\omega_{log}$, the contribution of H_DOS to the total DOS at the Fermi energy, $P_H$ (%), H-H bond length (Å), the charge obtained by each H atom (H_Charge), the average ICOHP value of the interaction between H atoms.

In order to understand the relationship between the structure and potential superconductivity of these quaternary hydrides, we calculated their $T_c$ values at 20 GPa using the Allen-Dynes modified Macmillan equation[36]:

$$T_c = \omega_{log} \frac{f_1 f_2}{1.2} \exp\left(\frac{-1.04(1+\lambda)}{\lambda - \mu^* - 0.62\lambda\mu^*}\right) \quad (1)$$

where $f_1$ and $f_2$ are two correction factors. The Coulomb pseudopotential $\mu^*$ is set to the typical 0.1-0.13. The $\omega_{log}$ and $\lambda$ were given by:

$$\omega_{log} = \exp\left(\frac{2}{\lambda} \int \frac{d\omega}{\omega} \alpha^2 F(\omega) \ln(\omega)\right) \quad (2)$$

$$\lambda = 2 \int \frac{\alpha^2 F(\omega)}{\omega} d\omega \quad (3)$$

Clearly, $T_c$ is determined by $\lambda$ and $\omega_{log}$, which are both related to $\alpha^2 F(\omega)$. As shown in the Figure 3, the $T_c$ values of the $XM_3Be_4H_{32}$ compounds basically increase with increasing values of lambda, although exhibiting non-monotonic variations as influenced by $\omega_{log}$. Strikingly, most of them host $T_c$ values above 100 K at 20 GPa, much higher than the liquid nitrogen temperature, especially the $AcLa_3Be_4H_{32}$ host has the highest $T_c$ value of 153 K.

As we know, the conventional superconductivity dominated by hydrogen in hydrides relies on the tuning of H-H bonds, which involves key factors such as the ionic radius, electronegativity, and valence of the metal atoms acting as pre-compressor charges. In general, the electrons donated by metal atoms can weaken the H-H covalent interaction and lengthen the H-H bond, but too long H-H bond will lead to the formation of isolated atom H, which will cause the hydrogen-related electronic state to be far away from the Fermi level. For example, in the A15-structured compound YZrH$_6$[42], each H atom can gain 0.57 electrons and the H-H bond length is 2.09 Å, which leads to a weak contribution of hydrogen to the density of electronic states at the Fermi level. Differently, in the XM$_3$Be$_4$H$_{32}$ structure, metal atoms with different stoichiometry regulate the number of electrons donated to hydrogen. Bader charge analysis shows that each H atom can gain about 0.41-0.45 electrons (see Table S2), these extra electrons partially occupy the H-H antibonding orbitals, preserving the H-H weak covalent interactions and thus driving the high H-DOS at the Fermi energy level. The calculated Crystal Orbital Hamilton Population (COHP) shows that all these quaternary hydrides exhibit significant H-H antibonding states at the Fermi energy level, which also occurs in previously reported caged superconducting hydrides, such as YH$_{10}$, LaH$_{10}$[4]. The integrated Crystal Orital Hamiltonian Population (ICOHP) values can well describe the bonding strength of the atom pair. The smaller -ICOHP value (0.18-0.25) again confirms the weak covalent interaction of the H-H bond (see Figure S8). The strong electron-phonon coupling of LaAc$_3$Be$_4$H$_{32}$ and AcLa$_3$Be$_4$H$_{32}$ is induced precisely by the significant H-derived electronic DOS at the Fermi energy level, thus hosting higher $T_c$ values. Ideally, for high $T_c$, the optimal dopant should maximize the electronic DOS at the Fermi level. However, the excessively high electronic DOS at the Fermi level may lead to severe softening of phonon vibration, resulting in lattice instability. Among the 42 quaternary hydrides, only 8 ones are dynamically stable. Their DOS of hydrogen at the Fermi energy level are in the range of about 1.25 to 1.75 (states/eV/f.u.), accounting for about 36 % of the total DOS (see Figure S9). These findings fully reveal that rationally tuning the synergistic relationship between the crystal structure, electronic structure, and electron-phonon coupling of quaternary hydrides, after which the expected stability and high-temperature superconductivity will be obtained.

We further investigated the critical pressure for the dynamic stability of these quaternary hydrides. The dynamic stable critical pressures of ThLa$_3$Be$_4$H$_{32}$, BaTh$_3$Be$_4$H$_{32}$ and AcTh$_3$Be$_4$H$_{32}$ are 10 GPa, and those of LaAc$_3$Be$_4$H$_{32}$ and AcLa$_3$Be$_4$H$_{32}$ are 20 GPa (see Figure S10-11). Remarkably, the dynamically stabilized critical pressures of 5 GPa for CaTh$_3$Be$_4$H$_{32}$, ThAc$_3$Be$_4$H$_{32}$, SrTh$_3$BeH$_{32}$ and LaTh$_3$Be$_4$H$_{32}$ are significantly lower than those of their parent structures (see Figure 4). The phonon branches in the low frequency region are mostly associated with the vibrations of metal atoms (La, Ca, Sr, Th, Ac) due to their heavier atomic masses, while the lighter Be and H atoms mainly contribute to the phonon modes in the mid-high frequency region. The softened phonon modes significantly enhance the electron-phonon coupling. For example, the soft phonon modes of the four structures shown in Figure 4 are mainly distributed in the range of 5-15 THz, such as LaTh$_3$Be$_4$H$_{32}$ near Γ point, SrTh$_3$Be$_4$H$_{32}$ and ThAc$_3$Be$_4$H$_{32}$ near M point and Γ point. The softening of the phonon modes of CaTh$_3$Be$_4$H$_{32}$ is very obvious, inducing the prominent peaks in the Eliashberg spectral function and boosting the electron-phonon coupling. In particular, the soft phonon modes in the 5-15 THz range contribute up to 48 % to the total electron-phonon coupling. Therefore, the chemical composition of multi-elements is diverse. In order to obtain strong electron-phonon coupling, the optimal combination of elements should allow for the softening of more phonon modes in the structure prior to dynamic instability, thus obtaining high-temperature superconductivity driven by strong electron-phonon coupling at moderate pressures. As in the case of CaTh$_3$Be$_4$H$_{32}$, which hosts the highest $T_c$ values among these four hydrides (124 K at 5 GPa).

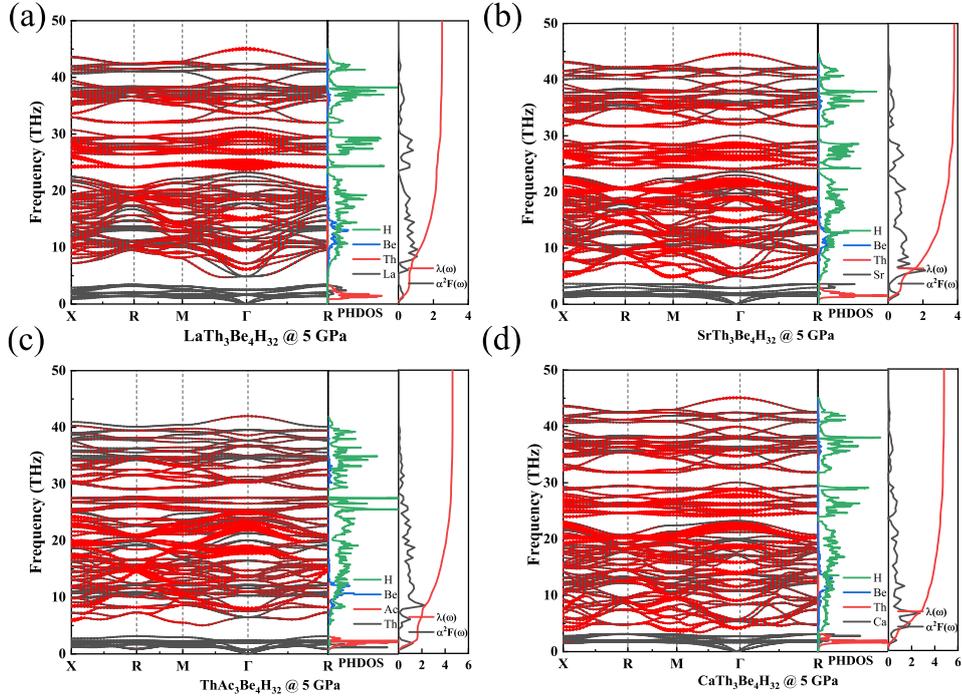

Fig.4 Calculated phonon dispersion curves, projected phonon density of states (PHDOS), and Eliashberg spectral function $a^2F(\omega)$ together with the electron-phonon integral $\lambda(\omega)$ for (a) LaTh$_3$Be$_4$H$_{32}$ (b) SrTh$_3$Be$_4$H$_{32}$ (c) ThAc$_3$Be$_4$H$_{32}$ (d) CaTh$_3$Be$_4$H$_{32}$ at 5 GPa. Red solid circles show the phonon linewidth with a radius proportional to the strength.

Next, let us focus on the pressure dependence of the structure and properties of these hydrides. In the case of CaTh$_3$Be$_4$H$_{32}$, as the pressure decreases, the number of electrons donated by the metal atoms to the H atoms gradually increases, lengthening the H-H bond and elevating the H-DOS at the Fermi energy level (see Figure 5). At the same time, lowering the pressure of superconducting hydrides especially when approaching dynamic instability, the softened lattice will enhance electron-phonon coupling, significantly boosting the $T_c$ value, but may encounter lattice dynamic instability. Figure S12 exhibits the region of distribution of the peaks of the spectral function moving progressively towards lower frequencies as the pressure decreases, and the peaks become larger, which is attributed to the significant softening of the phonon modes and the resulting significant effect on the total electron-phonon coupling. As expected, the $T_c$ value at 50 GPa is 100 K. As the pressure decreases, the EPC parameter gradually increases, especially at 5 GPa, the $\lambda$ value reaches 4.79, and the corresponding $T_c$ value is as high as 124 K.

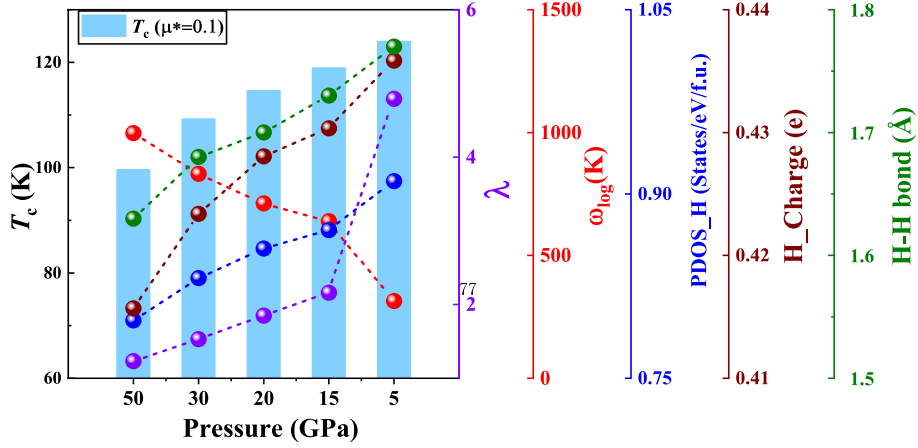

Fig. 5. The $T_c$ value and related parameters of $CaTh_3Be_4H_{32}$ as function of pressure. The $T_c$s are calculated using the Allen-Dynes modified McMillan equation ($\mu^* = 0.1$), the DOS of hydrogen at the Fermi energy (PDOS_H), the charge obtained by each H atom (H_Charge) and H-H bond length (Å).

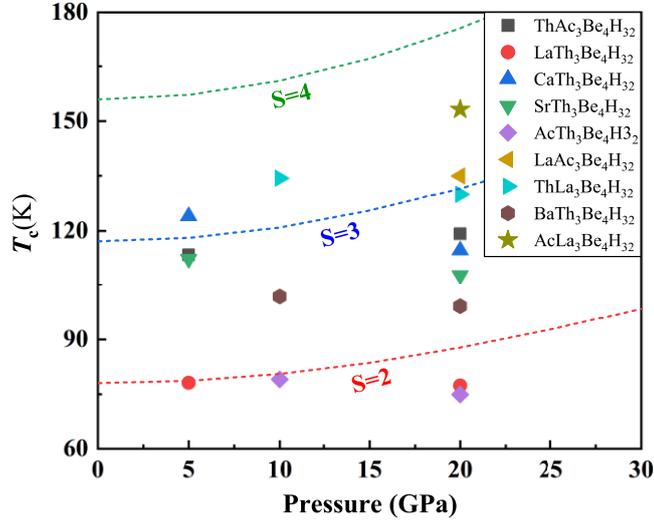

Fig. 6 The superconducting transition temperature of different $XM_3Be_4H_{32}$ compounds under pressure. The $T_c$s are calculated using the Allen-Dynes modified McMillan equation ($\mu^* = 0.1$). The background dotted lines are plotted based on the figure of merit S that assesses the significance of a particular superconductor ($S = T/\sqrt{P^2 + T_{MgB_2}^2}$)[21, 23].

In order to better evaluate the overall performance of superconductors, Pickard et al. defined the merit S [43], which is the result of the critical $T_c$ value and pressure balance of hydrides. Superconductors with significant advantages will show a large S value, that is, they tend to host higher $T_c$ values under lower pressure. As shown in the Figure 6, all these quaternary hydrides have high quality factors. The calculated S values of $ThLa_3Be_4H_{32}$, $CaTh_3Be_4H_{32}$, $LaAc_3Be_4H_{32}$ and $AcLa_3Be_4H_{32}$ exceed 3, especially the S value of $AcLa_3Be_4H_{32}$ reaches 3.5, which represents the highest value in quaternary hydrides. Thus, these quaternary hydrides are a class of superconductors with excellent comprehensive properties.

## 4. Conclusion

In summary, we investigated the structure and superconducting properties of the quaternary hydrides $XM_3Be_4H_{32}$ (X, M = Ca, Sr, Ba, Y, La, Ac, Th). Among the different chemical components considered, nine hydrides are dynamically stable at 20 GPa, with $CaTh_3Be_4H_{32}$, $ThAc_3Be_4H_{32}$, $SrTh_3BeH_{32}$ and $LaTh_3Be_4H_{32}$ as low as 5 GPa. Electron-phonon coupling calculations reveal that these compounds have higher $T_c$ than liquid nitrogen at mild pressures, and in particular $CaTh_3Be_4H_{32}$ (124 K at 5 GPa), $ThLa_3Be_4H_{32}$(134 K at 10 GPa), $LaAc_3Be_4H_{32}$ (135 K at 20 GPa) and $AcLa_3Be_4H_{32}$ (153 K at 20 GPa) host the high superconducting figure of merit S. The metal atoms with different ionic radii and electronegativities donate electrons to the hydrogen sublattice, thereby tuning the electronic structure near the Fermi level, especially inducing high H-DOS and strong Fermi surface nesting. The interactions between the high-energy vibrations associated with the hydrogen sublattice and the electrons from hydrogen induce strong electron-phonon coupling, giving rise to high-temperature superconductivity. These findings will help to deepen the understanding of the structure-property relationship of superconducting quaternary hydrides, more importantly, offer insights for exploring the hitherto still largely untapped multi-component superconducting hydrides, which hold great promise for exhibiting more diverse structural and superconducting characteristics.

**Declaration of Competing Interest**

The authors declare that they have no known competing financial interests or personal relationships that could have appeared to influence the work reported in this paper.

**Acknowledgements**

This work was supported by the National Natural Science Foundation of China (Grants No.

52072188, No. 12122405, No. 12274169), the National Key Research and Development Program of China (No. 2022YFA1402304), the Program for Science and Technology Innovation Team in Zhejiang (No. 2021R01004), and Jilin Provincial Science and Technology Development Project (No. 20210509038RQ). Some of the calculations were performed at the High-Performance Computing Center of Jilin University and using TianHe-1(A) at the National Supercomputer Center in Tianjin.

**References**

[1] D.F. Duan, Y.X. Liu, F.B. Tian, D. Li, X.L. Huang, Z.L. Zhao, H.Y. Yu, B.B. Liu, W.J. Tian, T. Cui, Pressure-induced metallization of dense $(H_2S)_2H_2$ with high-$T_c$ superconductivity, Sci. Rep. 4 (2014) 6968.

[2] A.P. Drozdov, M.I. Eremets, I.A. Troyan, V. Ksenofontov, S.I. Shylin, Conventional superconductivity at 203 kelvin at high pressures in the sulfur hydride system, Nature 525 (2015) 73-76.

[3] D.F. Duan, X.L. Huang, F.B. Tian, D. Li, H.Y. Yu, Y.X. Liu, Y.B. Ma, B.B. Liu, T. Cui, Pressure-induced decomposition of solid hydrogen sulfide, Phys. Rev. B 91 (2015) 180502.

[4] H.Y. Liu, Naumov, II, R. Hoffmann, N.W. Ashcroft, R.J. Hemley, Potential high-Tc superconducting lanthanum and yttrium hydrides at high pressure, Proc. Natl. Acad. Sci. USA 114 (2017) 6990-6995.

[5] F. Peng, Y. Sun, C.J. Pickard, R.J. Needs, Q. Wu, Y.M. Ma, Hydrogen clathrate structures in rare earth hydrides at high pressures: possible route to room-temperature superconductivity, Phys. Rev. Lett. 119 (2017) 107001.

[6] A.P. Drozdov, P.P. Kong, V.S. Minkov, S.P. Besedin, M.A. Kuzovnikov, S. Mozaffari, L. Balicas, F.F. Balakirev, D.E. Graf, V.B. Prakapenka, E. Greenberg, D.A. Knyazev, M. Tkacz, M.I. Eremets, Superconductivity at 250 K in lanthanum hydride under high pressures, Nature 569 (2019) 528-531.

[7] Z.M. Geball, H. Liu, A.K. Mishra, M. Ahart, Synthesis and stability of lanthanum superhydrides, Angew. Chem. Int. Ed. 57 (2018) 688-692.

[8] M. Somayazulu, M. Ahart, A.K. Mishra, Z.M. Geballe, M. Baldini, Y. Meng, V.V. Struzhkin, R.J. Hemley, Evidence for superconductivity above 260 K in lanthanum superhydride at megabar pressures, Phys. Rev. Lett. 122 (2019) 027001.

[9] F. Hong, L. Yang, P. Shan, P. Yang, Z. Liu, J. Sun, Y. Yin, X. Yu, J. Cheng, Z. Zhao, Superconductivity of lanthanum superhydride investigated using the standard four-probe configuration under high pressures, Chin. Phys. Lett. 37 (2020) 107401.

[10] I.A. Troyan, D.V. Semenok, A.G. Kvashnin, A.V. Sadakov, O.A. Sobolevskiy, V.M. Pudalov, A.G. Ivanova, V.B. Prakapenka, E. Greenberg, A.G. Gavriliuk, I.S. Lyubutin, V.V. Struzhkin, A. Bergara, I. Errea, R. Bianco, M. Calandra, F. Mauri, L. Monacelli, R. Akashi, A.R. Oganov, Anomalous high-temperature superconductivity in $YH_6$, Adv. Mater. 33 (2021) 2006832.

[11] P. Kong, V.S. Minkov, M.A. Kuzovnikov, A.P. Drozdov, S.P. Besedin, S. Mozaffari, L. Balicas, F.F. Balakirev, V.B. Prakapenka, S. Chariton, D.A. Knyazev, E. Greenberg, M.I. Eremets, Superconductivity up to 243 K in the yttrium-hydrogen system under high pressure, Nat. Commun. 12 (2021) 5075.

[12] E. Snider, N. Dasenbrock-Gammon, R. McBride, X. Wang, N. Meyers, K.V. Lawler, E. Zurek, A. Salamat, R.P. Dias, Synthesis of yttrium superhydride superconductor with a transition temperature

up to 262 K by catalytic hydrogenation at high pressures, Phys. Rev. Lett. 126 (2021) 117003.

[13] L. Ma, K. Wang, Y. Xie, X. Yang, Y. Wang, M. Zhou, H. Liu, X. Yu, Y. Zhao, H. Wang, G. Liu, Y. Ma, High-temperature superconducting phase in clathrate calcium hydride $CaH_6$ up to 215 K at a pressure of 172 GPa, Phys. Rev. Lett. 128 (2022) 167001.

[14] Y. Sun, J. Lv, Y. Xie, H. Liu, Y. Ma, Route to a superconducting phase above room temperature in electron-doped hydride compounds under high pressure, Phys. Rev. Lett. 123 (2019) 097001.

[15] D. An, D. Duan, Z. Zhang, Q. Jiang, H. Song, T. Cui, Thermodynamically stable room-temperature superconductors in Li-Na hydrides under high pressures, (2023) arXiv:2110.15628.

[16] M. Du, W. Zhao, T. Cui, D. Duan, Compressed superhydrides: the road to room temperature superconductivity, J. Phys.: Condens. Matter 34 (2022) 173001.

[17] D. Duan, Y. Liu, Y. Ma, Z. Shao, B. Liu, T. Cui, Structure and superconductivity of hydrides at high pressures, Natl. Sci. Rev. 4 (2017) 121-135.

[18] J.A. Flores-Livas, L. Boeri, A. Sanna, G. Profeta, R. Arita, M. Eremets, A perspective on conventional high-temperature superconductors at high pressure: Methods and materials, Phys. Rep. 856 (2020) 1-78.

[19] M.I. Eremets, A.P. Drozdov, P.P. Kong, H. Wang, Semimetallic molecular hydrogen at pressure above 350 GPa, Nat Phys 15 (2019) 1246-1249.

[20] W. Zhao, D. Duan, M. Du, X. Yao, Z. Huo, Q. Jiang, T. Cui, Pressure-induced high-$T_c$ superconductivity in the ternary clathrate system Y-Ca-H, Phys. Rev. B 106 (2022) 014521.

[21] M. Du, H. Song, Z. Zhang, D. Duan, T. Cui, Room-temperature superconductivity in Yb/Lu substituted clathrate hexahydrides under moderate pressure, Research 2022 (2022) 9784309.

[22] X. Feng, J. Zhang, G. Gao, H. Liu, H. Wang, Compressed sodalite-like $MgH_6$ as a potential high-temperature superconductor, RSC Adv. 5 (2015) 59292-59296.

[23] Z. Zhang, T. Cui, M.J. Hutcheon, A.M. Shipley, H. Song, M. Du, V.Z. Kresin, D. Duan, C.J. Pickard, Y. Yao, Design principles for high-temperature superconductors with a hydrogen-based alloy backbone at moderate pressure, Phys. Rev. Lett. 128 (2022) 047001.

[24] S. Di Cataldo, C. Heil, W. von der Linden, L. Boeri, $LaBH_8$: Towards high-$T_c$ low-pressure superconductivity in ternary superhydrides, Phys. Rev. B 104 (2021) L020511.

[25] X. Liang, A. Bergara, X. Wei, X. Song, L. Wang, R. Sun, H. Liu, R.J. Hemley, L. Wang, G. Gao, Y. Tian, Prediction of high-Tc superconductivity in ternary lanthanum borohydrides, Phys. Rev. B 104 (2021) 134501.

[26] S. Li, H. Wang, W. Sun, C. Lu, F. Peng, Superconductivity in compressed ternary alkaline boron hydrides, Phys. Rev. B 105 (2022) 224107.

[27] M. Gao, X.-W. Yan, Z.-Y. Lu, T. Xiang, Phonon-mediated high-temperature superconductivity in the ternary borohydride $KB_2H_8$ under pressure near 12 GPa, Phys. Rev. B 104 (2021) L100504.

[28] Y. Song, J. Bi, Y. Nakamoto, K. Shimizu, H. Liu, B. Zou, G. Liu, H. Wang, Y. Ma, Stoichiometric ternary superhydride mathrm LaBeH8 as a new template for high-temperature superconductivity at 110 K under 80 GPa, Phys. Rev. Lett. 130 (2023) 266001.

[29] G. Kresse, J. Furthmüller, Efficiency of ab-initio total energy calculations for metals and semiconductors using a plane-wave basis set, Comput. Mater. Sci. 6 (1996) 15-50.

[30] G. Kresse, D. Joubert, From ultrasoft pseudopotentials to the projector augmented-wave method, Phys. Rev. B 59 (1999) 1758-1775.

[31] P.E. Blöchl, Projector augmented-wave method, Phys. Rev. B 50 (1994) 17953-17979.

[32] J.P. Perdew, K. Burke, M. Ernzerhof, Generalized Gradient Approximation Made Simple Phys. Rev.

Lett. 77 (1996) 3865-3868.

[33] S. Maintz, V.L. Deringer, A.L. Tchougréeff, R. Dronskowski, LOBSTER: A tool to extract chemical bonding from plane-wave based DFT, J Comput Chem 37 (2016) 1030-1035.

[34] V.L. Deringer, A.L. Tchougréeff, R. Dronskowski, Crystal Orbital Hamilton Population (COHP) Analysis As Projected from Plane-Wave Basis Sets, J. Phys. Chem. A 115 (2011) 5461-5466.

[35] P. Giannozzi, S. Baroni, N. Bonini, M. Calandra, R. Car, C. Cavazzoni, D. Ceresoli, G.L. Chiarotti, M. Cococcioni, I. Dabo, A. Dal Corso, S. de Gironcoli, S. Fabris, G. Fratesi, R. Gebauer, U. Gerstmann, C. Gougoussis, A. Kokalj, M. Lazzeri, L. Martin-Samos, N. Marzari, F. Mauri, R. Mazzarello, S. Paolini, A. Pasquarello, L. Paulatto, C. Sbraccia, S. Scandolo, G. Sclauzero, A.P. Seitsonen, A. Smogunov, P. Umari, R.M. Wentzcovitch, QUANTUM ESPRESSO: a modular and open-source software project for quantum simulations of materials, J. Phys.: Condens. Matter 21 (2009) 395502.

[36] P.B. Allen, R.C. Dynes, Transition temperature of strong-coupled superconductors reanalyzed, Phys. Rev. B 12 (1975) 905-922.

[37] Y. Sun, S. Sun, X. Zhong, H. Liu, Prediction for high superconducting ternary hydrides below megabar pressure, J. Phys.: Condens. Matter 34 (2022) 505404.

[38] Q. Jiang, Z. Zhang, H. Song, Y. Ma, Y. Sun, M. Miao, T. Cui, D. Duan, Ternary superconducting hydrides stabilized via Th and Ce elements at mild pressures, Fundamental Research (2022).

[39] Z. Wan, R. Zhang, Metallization of hydrogen by intercalating ammonium ions in metal fcc lattices at lower pressure, Appl. Phys. Lett. 121 (2022) 192601.

[40] Y. Sun, M. Miao, Chemical templates that assemble the metal superhydrides, Chem 9 (2023) 443-459.

[41] J.S. Tse, Y. Yao, K. Tanaka, Novel Superconductivity in Metallic $\mathrm{SnH_4}$ under High Pressure, Phys. Rev. Lett. 98 (2007) 117004.

[42] W. Zhao, H. Song, M. Du, Q. Jiang, T. Ma, M. Xu, D. Duan, T. Cui, Pressure-induced high-temperature superconductivity in ternary Y–Zr–H compounds, Phys. Chem. Chem. Phys. 25 (2023) 5237-5243.

[43] C.J. Pickard, I. Errea, M.I. Eremets, Superconducting hydrides under pressure, Annu. Rev. Condens. Matter Phys. 11 (2020) 57-76.